# Experimental evidence of disordered crystalline premixing in sputter-deposited Ni(V)/Al multilayers

Michael J Abere,[1] Paul G. Kotula[1], Jonathan S. Paras,[1] and David P. Adams[1]

[1]Sandia National Laboratories, Albuquerque, NM USA 87123

**Abstract**

The sputter deposition of alternating layers of Ni(V) and Al forms a reactive multilayer known to undergo self-propagating formation reactions when ignited. The sequential deposition process leads to nm-scale premixing of reactants at each included interface which ultimately affects multilayer exothermicity. This work performs the direct measurement of a disordered face-centered cubic (FCC) solid solution premixed phase at the interfaces of Ni(V)/Al multilayers via scanning transmission electron microscopy. The crystallinity of the observed phase differs from previously reported a priori predictions of an amorphous interlayer. The disordered FCC phase retains its symmetry after annealing for 16 h at 135 ± 5°C, but the lattice parameter shifts consistent with an Al-rich composition. The existence of a crystalline premix in Ni(V)/Al is attributed to the electronic contribution to the entropy of crystallization.

Interfacial chemical reactions are a longstanding cornerstone of materials science, where short diffusion distances and large surface areas have enabled applications spanning microelectronics [1,2], microelectromechanical systems [3,4], and coatings [5]. In particular, interfacial structure underlies the performance of reactive nanomaterials composed of two or more reactants [6-10]. When subjected to an external heat source [18] or mechanical impulse [18,19], the stored chemical energy of a reactive nanomaterial can lead to a self-sustaining exothermic formation reaction. The magnitude of heat release involved with deflagration is generally less than the product formation enthalpy due to the existence of premixing at each interface, which is inherent to the material's fabrication process [20]. The relative thickness of premixed reactants at interfaces dictates requirements for ignition [21,22] and influences heat release rates [23], which are important to applications such as joining [11-13], brazing [14], sealing [15,16], and additive manufacturing [17]. In particular, the phase of the intermixed layer is decidedly important for shock loading, as it dictates properties like modulus, density, cohesive energy [24], and shock mixture models [25].

The phase of the intermixed region is often hedged as having the potential to exist in either solid solutions, intermetallic, or amorphous phases [26-28]. For reactive Ni/Al multilayers, the interfacial phase is commonly attributed an amorphous structure [22,29-32] because the standard practice of heating and quenching B2-NiAl within the LAMMPS molecular dynamics code with the accepted embedded atom potential [33] produces amorphous NiAl [30,31]. This is not a bad assumption, given that other sputter-deposited multilayers, Co/Al [34,35] and Al/Pt [36], have amorphous premixed layers. Furthermore, amorphous regions do form alongside nanocrystalline ones in ball-milled Ni + Al powders [37]. Providing insight into the intricate structure of reactive materials, this work examines the interfacial phase of vapor-deposited Ni(V)/Al multilayers. High-resolution scanning transmission electron microscopy (STEM) reveals a crystalline disordered solid solution NiAl (i.e., Ni and Al atoms randomly occupying a single FCC lattice) within the premixed regions of commercial sputter-deposited Ni(V)/Al foils.

Vapor-deposited Ni(V)/Al were purchased from Indium Corporation without the standard InCuSil™ or Sn capping layers. Energy dispersive spectroscopy (EDS) confirmed that the received product contained ~7 wt.% V within the Ni. The bilayer thickness was nominally 50 nm, with a total film thickness of 40 μm. A few foils were annealed in a CamCo 5 furnace that maintained a temperature of 135 ± 5°C to assess how interfacial structure changes with heat treatment. Cross-section STEM samples were prepared using focused ion beam lift-out in a FEI Co. Helios electron microscope. The STEM was performed with a 200 kV FEI Co. 80-200 aberration corrected Titan G2 microscope.

Cross-section bright field STEM of an as-received Ni(V)/Al multilayer and one annealed for 16 h at 135 ± 5°C are shown in Fig. 1. Both foils are oriented with the growth direction up. As subsequent layers are deposited, there is an observed increase in surface roughness and visible grain boundary grooving. A magnified bright field STEM image shows Ni-on-Al and Al-on-Ni interfaces in Fig. 2(a). In bright field contrast, Ni appears black, Al appears white, and the premixed material has an intermediate grayscale. There are visible lattice planes within the premix indicating a crystalline phase. A fast Fourier transform (FFT) of a single grain (the red box in Fig. 2(a)) aligned to the [100] zone axis produced reflections at 90° with a 1.88 Å spacing as shown in Fig. 2(b). Other grains produced reflections at 70° with a 2.17 Å spacing. These two reflections correspond to the (200) and (111) planes of an FCC (Fm3̄m) structure, respectively. Across 15 separate STEM images of the interface, no (100) reflections were observed, indicating that the mixed phase is composed of atoms on a single FCC Bravais lattice due to the lack of any primitive cell reflections (as dictated by the structure factor) from an ordered B2 or $L1_2$ phase. The observed pattern is consistent with a random site occupancy disordered phase. Direct measurements of the d-spacings for the (200) and (111) planes across 20 lattice planes within the premix are within 0.3% of those for a 50/50 disordered solid solution of $Ni_{0.93}V_{0.07}$ and Al. Moreover, the observed reflections would not be present in the monoclinic (P21/a) structure of potential $Al_9Ni_2$ or orthorhombic (Pnma) $Al_3Ni$ (i.e., the two lowest-temperature ordered intermetallics observed in Ni/Al multilayers) [20,38-41]. Thicknesses of the intermixed layer were determined by taking at least 20 line profiles of the bright field contrast across each interface and measuring the distance between the local maximum and minimum. For the as-received multilayer, there is an average of 2.15 nm (1.9 nm when Ni is grown on Al and 2.4 nm when Al is grown on Ni) premixed thickness.

The intermixed region from multilayers heated for 16 h at 135 ± 5°C, a temperature chosen below where pure Ni/Al multilayers show the formation of $Al_9Ni_2$ or $Al_3Ni$ in differential scanning calorimetry [38], was further characterized in STEM. Mirroring Fig. 2, a bright field image with FFT of the interfacial structure are provided in Fig. 3(a) and (b), respectively. The thermal cycle resulted in the growth of the intermixed layer thickness. Annealing it for 16 h produced a small growth of 1.60 nm to an average intermixed thickness of 3.75 nm (3.3 nm when Ni is grown on Al and 4.2 nm when Al is grown on Ni). The FFT in Fig. 3(b) shows a crystalline structure aligned to the [100] zone axis that retains an FCC structure with reflections from the (200) plane at 90°. However, the average d-spacing measures 1.98 Å, which is consistent with a more Al-rich disordered solid solution phase than the as-received premixing.

Concomitant EDS in Fig. 4 additionally shows the length over which the Al concentration declines from its ceiling to noise floor is on average 6.1 nm (6.4 nm when Ni is grown on Al and 5.8 nm when Al is grown on Ni) for the as-received and 8.3 nm (8.1 nm when Ni is grown on Al and 8.5 nm when Al is grown on Ni) for the 16 h annealed specimen — much larger than the distinct intermixed phase identified in the bright field images. The additional intermixed thickness contains Al concentrations beyond the equilibrium solubility limit [42] of either Ni in Al or Al in Ni, which may be due to the atomistic processes underlying multilayer fabrication by sputter deposition [36,43].

Understanding the crystallization behavior of thin film materials under vapor deposition processes necessitates a comprehensive examination of the competing effects of topology, thermodynamics, and electronic structure, all of which influence the tendency of materials to amorphize under conditions of high undercooling [44]. Historically, research in this area has concentrated on identifying alloys with a propensity to amorphize, guided by criteria such as those proposed by Inoue [45], which include the presence of three elemental components, an atomic size difference exceeding 15%, and large negative enthalpies of mixing among atomic constituents. Thermodynamic considerations have been developed that attempt to link the relative changes in ordering enthalpy and entropy, $\Delta H^{ord}$ and $\Delta S^{ord}$, to changing atomic bonding and system degrees of freedom [46]. These thermodynamic conditions, i.e., whether the $\Delta G^{amorphous} < \Delta G^{ordered}$, could be translated into kinetic predictions using Time-Temperature-Transformation diagrams to deduce the cooling rates necessary to achieve amorphization. As discussed in Wieczerzak's review [44], these criteria have yielded mixed results in systematically predicting glass formation in metallic systems.

The results presented here reveal a notable contrast to previous investigations of Co/Al premixing, which amorphized upon sputter deposition [34]. Conventional LAMMPS-supported molecular dynamics simulations failed to accurately predict the crystallization behavior of Ni/Al [30,31]. The difference in observed intermixed phase and LAMPSS predictions in Ni/Al originates from the fact that sputter deposition does not rapidly quench a liquid mixture of the two metals. This methodology results in amorphous premixed regions that match experimental measurements in both the Al/Pt and Co/Al bimetallic systems. There, LAMPSS accurately describes local variations in atomic complexions across sputter-deposited interfaces [47]. Despite both CoAl and NiAl exhibiting similar enthalpies of formation (-65.7 vs -61.8 kJ/g-atom) and crystal structures (ordered B2) [48,49], the differences in their amorphization behavior cannot be solely explained by enthalpic bonding considerations. Furthermore, neither material aligns with typical candidates for amorphous phase formation according to Inoue's selection rules. The implications of this suggest that significant changes in configuration and vibrational entropy are also unlikely, given their similar crystal symmetries and atomic weights. This leads to the consideration of the electron subsystems, which may play a substantial role in the order-disorder transformations observed in intermetallics [50].

Using a previously developed formalism [50], which allows for the calculation of equilibrium thermodynamic properties from electronic transport data, one can examine the conditions under which the electronic contribution to entropy may explain the system's propensity to amorphize. Experimental electronic transport data from Butler et al. [51] indicate that the thermopower (Seebeck coefficient [52]) of crystalline B2-CoAl is an order of magnitude larger and more negative than that of B2-NiAl thin films. This implies that the CoAl system exhibits a significantly larger electronic contribution to entropy, attributable to the greater density of states associated with a higher thermopower. We speculate that amorphization in systems with large densities of states, in accordance with Mott's theory on electronic structure as a function of order [53], may lead to even larger increases in state entropy compared to those systems lacking significant electronic contributions. This may explain why, under similar deposition conditions, Co/Al films amorphized at interfaces while their Ni/Al counterparts did not. Given that electronic effects are not considered in conventional molecular dynamics simulations, it is therefore unsurprising that such methods failed to predict the observed phase behavior.

In summary, this work characterizes the crystalline disordered solid solution within the premixed interfacial volumes of sputter-deposited Ni(V)/Al multilayers and underscores the unique nature of this system compared to other bimetallic systems, such as Co/Al and Al/Pt, which exhibit amorphous premixing. The difference in observed phase is attributed to differences in the electronic contribution to the entropy of crystallization of B2-NiAl, given its self-similarity to B2-CoAl in configurational entropy, vibrational entropy, and enthalpy of formation. While future work will require a deeper understanding of the various intermetallic Fermi surfaces to fully quantify differences in electronic entropy contributions, the observed crystalline premix in Ni/Al should be implemented into future reaction models throughout the energetic materials community henceforth.

**Acknowledgements**

The authors would like to thank Rob Knepper for his internal peer review. This work was supported by the Sandia National Laboratory Directed Research and Development (LDRD) program. Sandia National Laboratories is a multi-mission laboratory managed and operated by National Technology & Engineering Solutions of Sandia, LLC (NTESS), a wholly owned subsidiary of Honeywell International Inc., for the U.S. Department of Energy's National Nuclear Security Administration (DOE/NNSA) under contract DE-NA0003525. This written work is authored by an employee of NTESS. The employee, not NTESS, owns the right, title, and interest in and to the written work and is responsible for its contents. Any subjective views or opinions that might be expressed in the written work do not necessarily represent the views of the U.S. Government. The publisher acknowledges that the U.S. Government retains a non-exclusive, paid-up, irrevocable, world-wide license to publish or reproduce the published form of this written work or allow others to do so, for U.S. Government purposes. The DOE will provide public access to results of federally sponsored research in accordance with the DOE Public Access Plan.

References

[1] F.M. d'Heurle, Nucleation of a new phase from the interaction of two adjacent phases: Some silicides, J. Mater. Res. 3 (1988) 167–195. https://doi.org/10.1557/JMR.1988.0167.

[2] G. Levitin, D.W. Hess, Surface reactions in microelectronics process technology, Annu. Rev. Chem. Biomol. Eng. 2 (2011) 299–324. https://doi.org/10.1146/annurev-chembioeng-061010-114249.

[3] Y. Fu, H. Du, W. Huang, S. Zhang, M. Hu, TiNi-based thin films in MEMS applications: a review, Sensors Actuators A: Phys. 112 (2004) 395-408. https://doi.org/10.1016/j.sna.2004.02.019

[4] M. Guo, J.T. Brewster II, H. Zhang, Y. Zhao, Y. Zhao, Challenges and opportunities of chemiresistors based on microelectromechanical systems for chemical olfaction, ACS Nano 16 (2022) 17778–17801. https://doi.org/10.1021/acsnano.2c08650.

[5] A. Ghadi, H. Ebrahimnezhad-Khaljiri, R. Gholizadeh, A comprehensive review on the carbide-base coatings produced by thermo-reactive diffusion: microstructure and properties viewpoints, J. Alloys Compd. (2023) 171839. https://doi.org/10.1016/j.jallcom.2023.171839.

[6] M.M.P. Janssen, G.D. Rieck, Reaction diffusion and Kirkendall-effect in the nickel-aluminum system, Trans. Metall. Soc. AIME 239 (1967) 1372–1385.

[7] E.G. Colgan, A review of thin-film aluminide formation, Mater. Sci. Rep. 5 (1990) 1–44. https://doi.org/10.1016/S0920-2307(05)80005-2.

[8] P. Zhu, J.C.M. Li, C.T. Liu, Combustion reaction in multilayered nickel and aluminum foils, Mater. Sci. Eng. A 239 (1997) 532–539. https://doi.org/10.1016/S0921-5093(97)00627-8.

[9] P. Zhu, J.C.M. Li, C.T. Liu, Reaction mechanism of combustion synthesis of NiAl, Mater. Sci. Eng. A 329 (2002) 57–68. https://doi.org/10.1016/S0921-5093(01)01549-0.

[10] J. Zhang, F. Zhao, H. Li, Z. Yuan, M. Zhang, Y. Yang, ... Z. Qin, Improving ignition and combustion performance of Al@ Ni in CMDB Propellants: Effect of nickel coating, Chem. Eng. J. 456 (2023) 141010. https://doi.org/10.1016/j.cej.2022.141010.

[11] J. Wang, E. Besnoin, A. Duckham, S.J. Spey, M.E. Reiss, O.M. Knio, ... T.P. Weihs, Room-temperature soldering with nanostructured foils, Appl. Phys. Lett. 83 (2003) 3987–3989. https://doi.org/10.1063/1.1623943.

[12] X. Cai, X. Ren, C. Sang, L. Zhu, Z. Li, P. Feng, Dissimilar joining mechanism, microstructure and properties of Ni to 316 stainless steel via Ni-Al thermal explosion reaction, Mater. Sci. Eng. A 807 (2021) 140868. https://doi.org/10.1016/j.msea.2021.140868.

[13] M. Glaser, S. Matthes, J. Hildebrand, J.P. Bergmann, P. Schaaf, Hybrid Thermoplastic-Metal joining based on Al/Ni multilayer foils–Analysis of the joining zone, Mater. Des. 226 (2022) 111561. https://doi.org/10.1016/j.matdes.2022.111561.

[14] C. Suryanarayana, J.J. Moore, R.P. Radtke, Novel methods of brazing dissimilar materials, Adv. Mater. Process. 159 (2001) 29.

[15] J. Braeuer, J. Besser, M. Wiemer, T. Geßner, A novel technique for MEMS packaging: Reactive bonding with integrated material systems, Sensors Actuators A: Phys. 188 (2012) 212–219. https://doi.org/10.1016/j.sna.2012.01.015.

[16] M. Powers, J. Subramanian, J. Levin, T. Rude, D. Van Heerden, O. Knio, Room temperature hermetic sealing of microelectronic packages with nanoscale multilayer reactive foils, Mater. Sci. Technol. Assoc. Iron Steel Technol. 3 (2005) 2.

[17] R. Liu, C. Gao, A. Hou, S. Wang, Ni/Al foil-based reactive additive manufacturing with fast rate and high energy-efficiency, J. Mater. Process. Technol. 321 (2023) 118167.

[18] G.M. Fritz, et al., Thresholds for igniting exothermic reactions in Al/Ni multilayers using pulses of electrical, mechanical, and thermal energy, J. Appl. Phys. 113 (2013). https://doi.org/10.1063/1.4770478.

[19] S.C. Kelly, N.N. Thadhani, Shock compression response of highly reactive Ni+ Al multilayered thin foils, J. Appl. Phys. 119 (2016). https://doi.org/10.1063/1.4942931.

[20] A.J. Gavens, D. Van Heerden, A.B. Mann, M.E. Reiss, T.P. Weihs, Effect of Intermixing on Self-Propagating Exothermic Reactions in Al/Ni Nanolaminate Foils, J. Appl. Phys. 87 (2000) 1255–1263. https://doi.org/10.1063/1.372005.

[21] A. Hémeryck, J.M. Ducéré, C. Lanthony, A. Estève, C. Rossi, M. Djafari-Rouhani, D. Estève, Bottom-up modeling of Al/Ni multilayer combustion: Effect of intermixing and role of vacancy defects on the ignition process, J. Appl. Phys. 113 (2013) 20. https://doi.org/10.1063/1.4807164.

[22] Y. Xie, K. Zhu, F. Xue, J.L. Shao, P. Chen, Premixing degree-dependent reaction mechanisms of premixed Ni/Al nanolaminates under shock loading, Phys. Fluids 36 (2024) 8. https://doi.org/10.1063/5.0216018.

[23] F. Schwarz, R. Spolenak, The influence of premixed interlayers on the reaction propagation in Al–Ni multilayers—An MD approach, J. Appl. Phys. 131 (2022) 7. https://doi.org/10.1063/5.0079035.

[24] L. Sandoval, G.H. Campbell, J. Marian, Thermodynamic interpretation of reactive processes in Ni–Al nanolayers from atomistic simulations, Modelling Simul. Mater. Sci. Eng. 22 (2014) 2. https://doi.org/10.1063/5.0237889.

[25] D.E. Kittell, P.E. Specht, M.J. Abere, D.P. Adams, Continuum Shock Mixture Models for Ni/Al Multilayers: Individual Layers and Bulk Equations of State, J. Appl. Phys. 2025. https://doi.org/10.1063/5.0171539.

[26] S. Sen, M. Lake, P. Schaaf, Al-based binary reactive multilayer films: Large area freestanding film synthesis and self-propagating reaction analysis, Appl. Surf. Sci. 474 (2019) 243–249. https://doi.org/10.1016/j.apsusc.2018.02.207.

[27] A.S. Ramos, S. Simões, L. Maj, J. Morgiel, M.T. Vieira, Effect of deposition parameters on the reactivity of Al/Ni multilayer thin films, Coatings 10 (2020) 8. https://doi.org/10.3390/coatings10080721.

[28] N. Toncich, F. Schwarz, R.A. Gallivan, J. Gillon, R. Spolenak, Composition Effects on Ni/Al Reactive Multilayers: A Comprehensive Study of Mechanical Properties, Reaction Dynamics and Phase Evolution, arXiv preprint arXiv:2502.02333 (2025). https://doi.org/10.1063/5.0263283.

[29] V. Turlo, O. Politano, F. Baras, Modeling self-sustaining waves of exothermic dissolution in nanometric Ni-Al multilayers, Acta Mater. 120 (2016) 189–204. https://doi.org/10.1016/j.actamat.2016.08.014.

[30] Y. Xie, J.L. Shao, R. Liu, P. Chen, Atomic insights into shock-induced alloying reaction of premixed Ni/Al nanolaminates, J. Chem. Phys. 159 (2023) 17. https://doi.org/10.1063/5.0171468.

[31] F. Schwarz, R. Spolenak, The influence of premixed interlayers on the reaction propagation in Al–Ni multilayers—An MD approach, J. Appl. Phys. 131 (2022) 7. https://doi.org/10.1063/5.0079035.

[32] A. Fourmont, O. Politano, S. Le Gallet, C. Desgranges, F. Baras, Reactivity of Ni–Al nanocomposites prepared by mechanical activation: A molecular dynamics study, J. Appl. Phys. 129 (2021) 6. https://doi.org/10.1063/5.0037397.

[33] G.P. Purja Pun, Y. Mishin, Development of an interatomic potential for the Ni-Al system, Philos. Mag. 89 (2009) 34-36, 3245–3267. https://doi.org/10.1080/14786430903258184.

[34] D.P. Adams, V.C. Hodges, M.M. Bai, E. Jones, M.A. Rodriguez, T. Buchheit, J.J. Moore, Exothermic Reactions in Co/Al Nanolaminates, J. Appl. Phys. 104 (2008) 043502. https://doi.org/10.1063/1.2968444.

[35] M.J. Abere, R.V. Reeves, C. Sobczak, H. Choi, P.G. Kotula, D.P. Adams, Effects of diffusion barriers on reaction wave stability in Co/Al reactive multilayers, J. Appl. Phys. 134 (2023) 19. https://doi.org/10.1063/5.0171539.

[36] D.P. Adams, R.V. Reeves, M.J. Abere, C. Sobczak, C.D. Yarrington, M.A. Rodriguez, P.G. Kotula, Ignition and self-propagating reactions in Al/Pt multilayers of varied design, J. Appl. Phys. 124 (2018) 9. https://doi.org/10.1063/1.5026293.

[37] A.S. Rogachev, N.F. Shkodich, S.G. Vadchenko, F. Baras, D.Y. Kovalev, S. Rouvimov, A.A. Nepapushev, A.S. Mukasyan, Influence of the high energy ball milling on structure and reactivity of the Ni+ Al powder mixture, J. Alloys Compd. 577 (2013) 600–605. https://doi.org/10.1016/j.jallcom.2013.06.114.

[38] K.J. Blobaum, D. Van Heerden, A.J. Gavens, T.P. Weihs, Al/Ni Formation Reactions: Characterization of the Metastable Al9Ni2 Metastable Phase and Analysis of Its Formation, Acta Mater. 51 (2003) 3871–3884. https://doi.org/10.1016/S1359-6454(03)00211-8.

[39] H. Nathani, J. Wang, T.P. Weihs, Long-term Stability of Nanostructured Systems with Negative Heats of mixing, J. Appl. Phys. 101 (2007) 104315. https://doi.org/10.1063/1.2736937.

[40] M.D. Grapes, T. LaGrange, K. Woll, B.W. Reed, G.H. Campbell, D.A. LaVan, T.P. Weihs, In situ transmission electron microscopy investigation of the interfacial reaction between Ni and Al during rapid heating in a nanocalorimeter, APL Mater. 2 (2014) 116102. https://doi.org/10.1063/1.4900818.

[41] A.S. Rogachev, S.G. Vadchenko, F. Baras, O. Politano, S. Rouvimov, N.V. Sachkova, M.D. Grapes, T.P. Weihs, A.S. Mukasyan, Combustion in reactive multilayer Ni/Al nanofoils: Experiments and molecular dynamic simulation, Combust. Flame 166 (2016) 158–169. https://doi.org/10.1016/j.combustflame.2016.01.014.

[42] I. Ansara, N. Dupin, H.L. Lukas, B. Sundman, Thermodynamic assessment of the Al-Ni system, J. Alloys Compd. 247 (1997) 1-2, 20–30. https://doi.org/10.1016/S0925-8388(96)02652-7.

[43] M.J. Abere, G.C. Egan, D.E. Kittell, G.H. Campbell, D.P. Adams, Phase mediated dynamics of self-propagating Co/Al nanolaminate reactions, AIP Adv. 10 (2020) 8. https://doi.org/10.1063/5.0015317.

[44] K. Wieczerzak, et al., Crystalline or amorphous? A critical evaluation of phenomenological phase selection rules, Mater. Des. 230 (2023) 111994. https://doi.org/10.1016/j.matdes.2023.111994.

[45] A. Inoue, Stabilization of metallic supercooled liquid and bulk amorphous alloys, Acta Mater. 48 (2000) 1, 279–306. https://doi.org/10.1016/S1359-6454(99)00300-6.

[46] A.K. Singh, et al., A geometrical parameter for the formation of disordered solid solutions in multi-component alloys, Intermetallics 53 (2014) 112–119. https://doi.org/10.1016/j.intermet.2014.04.019.

[47] C.B. Saltonstall, Z.D. McClure, M.J. Abere, D. Guzman, S.T. Reeve, A. Strachan, P.G. Kotula, D.P. Adams, T.E. Beechem, Complexion dictated thermal resistance with interface density in reactive metal multilayers, Phys. Rev. B 101 (2020) 24, 245422. https://doi.org/10.1103/PhysRevB.101.245422.

[48] P. Nash, O. Kleppa, Composition dependence of the enthalpies of formation of NiAl, J. Alloys Compd. 321 (2001) 2, 228–231. https://doi.org/10.1016/S0925-8388(01)00952-5.

[49] E.-T. Henig, H.L. Lukas, G. Petzow, Enthalpy of Formation and Description of the Defect Structure of the Ordered ß-Phase in Co-Al: Dedicated to Prof. Dr. Konrad Schubert on his 65th birthday, Int. J. Mater. Res. 71 (1980) 6, 398–402.

[50] J. Paras, A. Allanore, Contribution of electronic entropy to the order-disorder transition of $Cu_3Au$, Phys. Rev. Res. 3 (2021) 2, 023239. https://doi.org/10.1103/PhysRevResearch.3.023239.

[51] S.R. Butler, J.E. Hanlon, R.J. Wasilewski, Electric and magnetic properties of B2 structure compounds: NiAl, CoAl, J. Phys. Chem. Solids 30 (1969) 8, 1929–1934. https://doi.org/10.1016/0022-3697(69)90168-1.

[52] S. *Blundell, S.J. Blundell, K.M. Blundell (2010). Concepts in Thermal Physics. Oxford University Press. p. 415.* ISBN 978-0-19-956210-7.

[53] N.F. Mott, Electrons in disordered structures, Adv. Phys. 16 (1967) 61, 49-144. https://doi.org/10.1080/00018736700101265.

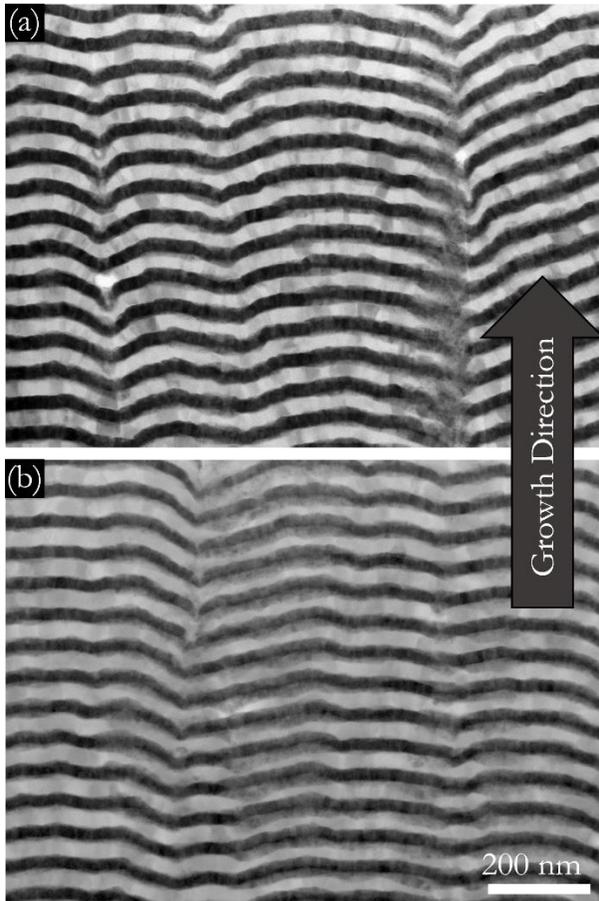

**Figure 1.** Bright field STEM image of an (a) as-received and (b) 16 hours annealed at 135 ± 5°C commercial 50 nm bilayer Ni/Al multilayer.

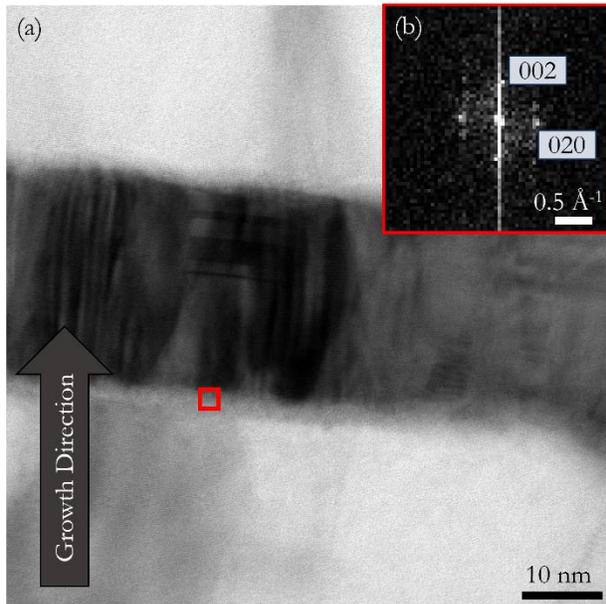

**Figure 2.** (a) Bright field STEM image of an as-received commercial 50 nm bilayer Ni/Al multilayer. (b) Fast Fourier transform of a nanocrystalline grain at the interface showing the [100] zone axis of a disordered FCC structure.

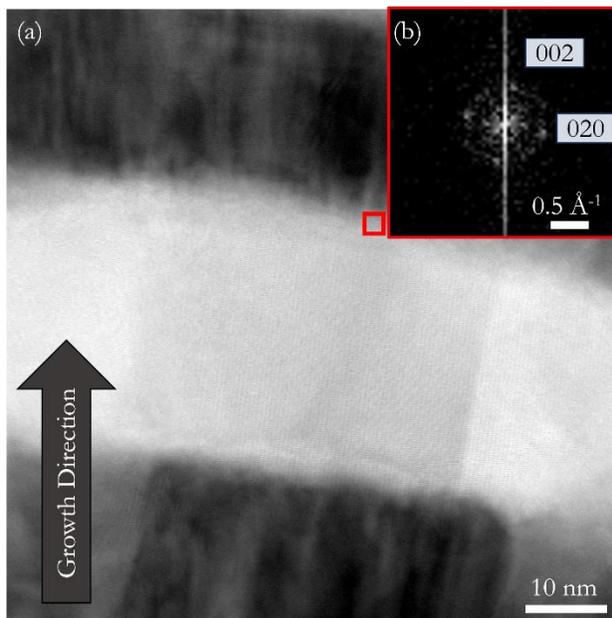

**Figure 3.** (a) Bright field STEM image of a commercial 50 nm bilayer Ni/Al multilayer annealed for 16 hours at 135 ± 5°C. (b) Fast Fourier Transform of a nanocrystalline grain at the interface showing the [100] zone axis of a disordered FCC structure.

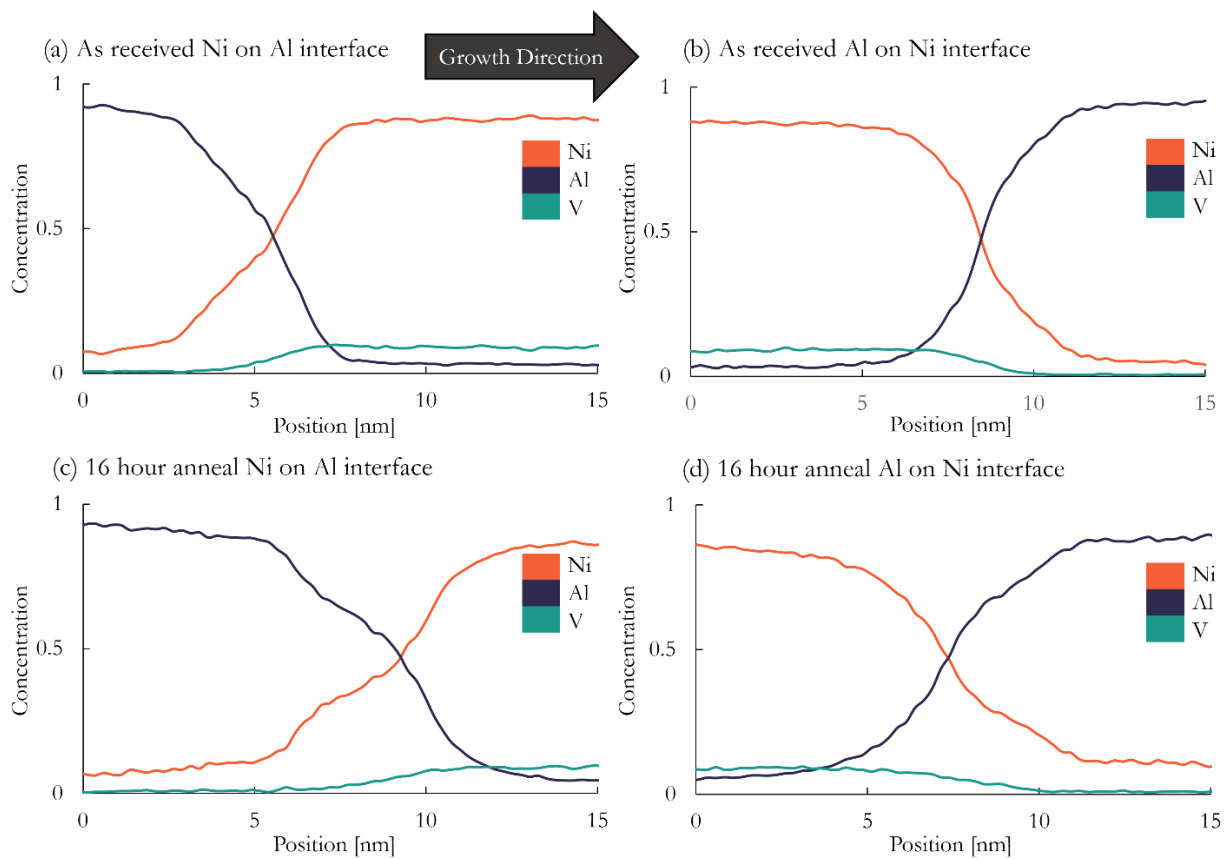

**Figure 4.** (a) and (b) EDS line plot of the interfacial composition of as-received commercial 50 nm bilayer Ni/Al multilayer. (c) and (d) EDS line plot showing the change to interfacial composition after annealing for 16 hours at 135 ± 5°C.